\documentclass[
    aip,
    amsmath,
    amssymb,
    reprint,
]{revtex4-1}

\usepackage[utf8]{inputenc}
\usepackage[T1]{fontenc}
\usepackage[colorlinks=true,citecolor=blue,allcolors=blue]{hyperref}
\usepackage{amsmath,amsthm,amsfonts,amssymb,amscd}
\usepackage{graphicx}
\usepackage{dcolumn}
\usepackage{enumerate}
\usepackage{bm}
\usepackage{mathptmx}
\usepackage{braket}
\usepackage[version=4]{mhchem}

\newcommand{\quotes}[1]{``#1''}

\draft 

\begin{document}

\title[Bound and autoionizing potential energy curves in the CH molecule]{Bound and autoionizing potential energy curves in the CH molecule}

\author{D\'avid Hvizdo\v{s}}
\email{hvizdosdavid@gmail.com}
\affiliation{Department of Physics and Astronomy and Purdue Quantum Science and Engineering Institute, Purdue University, West Lafayette, Indiana, 47907, USA}
\author{Joshua Forer}
\affiliation{Department of Physics, University of Central Florida, Florida, 32816, USA}
\affiliation{Institut des Sciences Moléculaires, Université de Bordeaux, CNRS UMR 5255, 33405, Talence Cedex, France}
\author{Viatcheslav Kokoouline}
\affiliation{Department of Physics, University of Central Florida, Florida, 32816, USA}
\author{Chris H. Greene}
\affiliation{Department of Physics and Astronomy and Purdue Quantum Science and Engineering Institute, Purdue University, West Lafayette, Indiana, 47907, USA}

\date{\today}

\begin{abstract}
This article presents a method of computing bound state potential curves and autoionizing curves using fixed-nuclei R-matrix data extracted from the Quantemol-N software suite. It is a method based on two related approaches of multichannel quantum-defect theory. One is applying bound-state boundary conditions to closed-channel asymptotic solution matrices and the other is searching for resonance positions via eigenphase shift analysis. We apply the method to the CH molecule to produce dense potential-curve data sets presented as graphs and supplied as tables in the publication supplement.
\end{abstract}


\maketitle 

\section{\label{sec-intro}Introduction}

The CH molecule and its positive ion CH$^+$ play an important role in certain molecular plasma environments, such as in the interstellar medium or electric discharges and flames of hydrocarbon gases. The process, dominant in destruction of the CH$^+$ ions in such environments, is typically dissociative recombination (DR) with electrons: The ion recombines with an incident electron, forming a neutral molecule, typically, in a highly-excited electronic state of CH, and then the molecule dissociates. Our recent work \cite{CHplus_DR} was devoted to the process. It was found that the three lowest electronic states (X$^1\Sigma^+$, a$^3\Pi$, and A$^1\Pi$) of the  CH$^+$ ion contribute to the process: they produce low-energy resonances in the DR spectrum, increasing significantly the DR cross section. The presence of multiple low-energy excited states means that the resulting CH potential and autoionizing curves contain multiple crossing series of Rydberg states, which result in many complex avoided crossings. This study is devoted to the determination of highly-excited electronic bound states of CH and autoionizing resonances.

Motivated, in particular, by the DR process, the lowest electronic states of the CH molecule have been studied in many theoretical and experimental studies. Here, we give a short overview of theoretical studies.  As early as in 1970, \citet{liu1970electronic} have used the LCAO-MO-SCF method and computed potential energy curves (PEC) as functions of the internuclear distance for nine low electronic states. Later,  \citet{bardsley1973dissociative}, also \citet{krauss1973dissociative} have studied in more detail energies of the CH electronic states near  the equilibrium position and energy of the ground state of the ion (X$^1\Sigma^+$) in order to assess how fast the DR process is in CH$^+$. They have used a method \cite{PhysRev.135.A1227} of mixed orbital sets composed of  Slater-type orbitals and elliptic orbitals. In 1977,  \citet{giusti1977dissociative} have also performed calculations of several PECs at energies and geometries near the equilibrium of the ion, using the ion virtual orbital method of
Lefebvre-Brion \cite{lefebvre1973calculation}. For a photodissociation study,
\citet{van1987photodissociation} determined PECs for more than a dozen of electronic states of CH using a method of multireference configuration interaction (MRCI). Interestingly, knowing now quite well the CH and CH$^+$ structure at low energies, we can notice that the region of energies near the CH$^+$ equilibrium in the study by \citet{van1987photodissociation} was not computed as accurately as in earlier studies, such as by \citet{giusti1977dissociative}. Soon later, \citet{takagi1991dissociative}, for a study on DR in CH$^+$, have computed several PECs using the ALCHEMY code based on the Slater-type orbitals, obtaining very accurate results, at least, near the  CH$^+$ equilibrium. In 1999 \citet{kalemos99} have published PECs also for more than a dozen of CH electronic states, obtained using a MRCI method and the Molpro suite of codes. More recently,  \citet{vazquez2007insight} have published very accurate PECs of the lowest states obtained also with MRCI. The focus of that study  was  3$d$ Rydberg states of CH, assigned previously to be responsible for two observed but unidentified interstellar bands.  We mention here also spectroscopy-accurate  coupled-cluster calculations by \citet{shi2008accurate} of the PEC of the ground electronic state of CH. In a more recent calculation, \citet{chakrabarti2019r} have employed the UK R-matrix code \cite{quantemol3} to produce a few PEC of the Rydberg states, as well as energies and widths of the lowest autoionizing states of CH. The approach employed in that study is very different from the quantum-chemistry approaches used in other studies mentioned above.

The theoretical approach presented here is very different from all methods, mentioned above, while it has some similarity to the one used by  \citet{chakrabarti2019r}: It employs also the R-matrix formalism. It is more  similar in spirit to the method used by \citet{Little_2013} in their calculations of PECs of N$_2$, which employs quantum defect theory \cite{Seaton_2002,Aymar_Greene_LKoenig_1996} and certain elements of the scattering theory even for bound state calculations.
There are advantages to working with the scattering information in the smoothest energy- and $R$-dependences possible, especially if a rovibrational frame transformation will be ultimately carried out, e.g., to determine inelastic scattering or rearrangement cross sections such as dissociative recombination\cite{CHplus_DR}.  For this reason, we change from the R-matrix parameterization of the scattering data to a multichannel quantum defect theory (MQDT) representation (i.e. the $K$-matrix or its equivalent representation in terms of eigenquantum defects and eigenvectors) with comparatively smooth energy dependences.  There is a slight cost in accuracy by doing this, namely the fact that by matching to Coulomb functions at the R-matrix boundary $r_0$, we neglect the long range couplings at $r>r_0$.  Our tests show that those couplings are rather small in practice, and can be approximately treated perturbatively when higher accuracy is desirable.

Unless otherwise stated, we use atomic units throughout this article.

\section{\label{sec-theory}Theoretical outline}

The main part of this method is in applying asymptotic MQDT procedures to fixed-nuclei electron-scattering R matrix data.
We obtain our \ce{CH+ + e-} electron-scattering R matrix by employing the Quantemol-N software suite \cite{quantemol1} which uses the UKRMol \cite{quantemol2,quantemol3,quantemol4} R-matrix method. The main output of this software is the energy-dependent reactance matrix K at a large electronic distance, with its elements representing physically open channels. The relevant quantum numbers of these channels are the electronic state of the target ion $n$, the orbital angular momentum $l$ of the scattering electron and its projection $\lambda$ onto the molecular axis (though $\lambda$ will not explicitly appear in our later derivations). For simplicity, let us combine these into a joint index $i=\{n_i,l_i,\lambda_i\}$. Let us also denote the channel threshold $E_i$ and the channel energies $\epsilon_i=E-E_i$, where $E$ is some given total energy in Hartree energy units. It will also be useful to define the effective quantum number  for closed channels
\begin{equation}
    \label{eq-nu}
    \nu_i = (-2\epsilon_i)^{-1/2}.
\end{equation}
These Quantemol calculations employ the configuration-interaction method (CI). The correlation consistent polarized valence quadruple-zeta (cc-pVQZ) basis set was used to represent the electronic state of the target ion.
In the CI expansion, the lowest orbital $1\sigma^2$ was frozen, while the remaining four electrons were distributed over the  $2-7\sigma, 1-3\pi, 1\delta$ orbitals of the active space. Three $\sigma$ and two $\pi$ virtual orbitals were used to augment the continuum orbital set in the inner region scattering wave function.
The three lowest target states X$^1\Sigma^+$, a$^3\Pi$, and A$^1\Pi$ were retained in the closed-coupling expansion in the outer region.
The orbital angular momentum of the scattered electron goes up to 4 and the R-matrix box radius $r_0$ was 13 bohr radii. The coordinate system places the CH$^+$ center of mass at its origin.

\subsection{\label{sec-theory-RK} Extracting the R and K matrices}

An important intermediate output of Quantemol is the fixed-nuclei R-matrix data. The program solves the Schr\"odinger equation inside a restricted electronic box with an arbitrary boundary condition. The solutions $\ket{\psi_p}$ are then projected onto physical outer-region channels $\ket{i}$ to get the surface amplitudes $w_{ip} =\left\langle\left\langle i|\psi_p \right\rangle\right\rangle$. 
The double angle bracket indicates that the integration is only carried out on the surface of a sphere of radius $r_0$ centered about the center of mass of the \ce{CH+} ion, as opposed to integration over all space.
Note that these quantities are energy-independent and at this point it cannot be said whether an outer-region channel $i$ is physically open or closed. The saved surface amplitudes $w_{ip}$ along with the inner-solution R-matrix pole energies $E_p$ can be plugged into the Wigner-Eisenbud \cite{Wigner_Eisenbud_1947} formula to give the R matrix at an arbitrary energy:

\begin{equation}
    \label{eq-Rmat-WE}
    R_{ii'} = \frac{1}{2}\sum_p
    \frac{
        w_{ip} \, w^*_{i'p}}
        {E_p - E}\; .
\end{equation}

The short-range R matrix obtained via Eq. (\ref{eq-Rmat-WE}) has indices of both open and closed channels. If we wish, we can transform it using a preliminary elimination procedure (closely related to the MQDT closed-channel elimination procedure), which neglects long-range multipole couplings at $r>r_0$:
\begin{equation}
    \label{eq-chan-elim-R}
    \underline{R}^\mathrm{phys} = \underline{R_{oo}} - \underline{R_{oc}}
    \left[ \underline{R_{cc}} - \underline{W_c}\left(\underline{W'_c}\right)^{-1} \right]^{-1} \underline{R_{co}}\; ,
\end{equation}
where $o$ and $c$ denote blocks of open and closed channels respectively and $W_c$ is a diagonal matrix of the asymptotically decaying (for negative energies) Whittaker functions evaluated at the closed-channel energies, using the R-matrix box radius and the respective electronic angular momenta. $W'_c$ is its derivative with respect to the electronic coordinate. In our derivation, we use the "energy-normalized" Whittaker functions\cite{Aymar_Greene_LKoenig_1996}, but the division $W_c/W'_c$ makes re-scaling by any $r$-independent constant irrelevant in implementation. We can select which closed channels we eliminate this way (e.g. we can keep weakly closed channels while eliminating the strongly closed ones). For the derivation of this formula, see the appendix. A note for readers less familiar with MQDT: this elimination of channels does not mean erasing physical information but merely decreasing the dimension of the R matrix by applying asymptotic boundary conditions.

Upon obtaining the R matrix we transform it into the K matrix using the well-known relation \cite{Aymar_Greene_LKoenig_1996}
\begin{equation}
\label{eq-R2K}
    \underline{K} = \left( \underline{f} -
    \underline{f}' \underline{R} \right)
    \left( \underline{g} -
    \underline{g}' \underline{R} \right)^{-1}\; ,
\end{equation}
where $\underline{f}$ and $\underline{g}$ are diagonal matrices of the regular and irregular Coulomb functions $f(\epsilon_i,l_i,r_0)$, $g(\epsilon_i,l_i,r_0)$.
Primed quantities here and throughout denote an element-wise derivative with respect to the electronic coordinate, $r$.
The pair $\{f,g\}$ is equal to the pair $\{s,-c\}$ in Seaton's work\cite{Seaton_2002}. These functions are well defined for all open channels and for weakly closed channels as long as $\epsilon_i > -1/(2l_i^2)$. For lower energies, the functions become complex. Thus, for closed channels with high $l$ values ($l\geq2$) that we do not eliminate, we replaced the pair $\{f,g\}$ with the rescaled $\{f^{\eta},g^{\eta}\}$,
\begin{align}
\label{eq-F-rescale}
    f^{\eta}(\epsilon,l,r) &= f(\epsilon,l,r)/\sqrt{A(\epsilon,l)}, \\
\label{eq-G-rescale}
    g^{\eta}(\epsilon,l,r) &= g(\epsilon,l,r) \sqrt{A(\epsilon,l)},
\end{align}
where $A(\epsilon,l)=\prod_{j=0}^l(1+2j^2\epsilon)$. This alternate pair has also been used in previous works\cite{Greene_Fano_strinati,Jungen_Telmini}, though often rescaled by a factor of $\sqrt{2}$. This choice removes the complex behavior at negative energies but also alters some of the equations in the following sections.

The final step before branching into either the potential curve or autoionizing curve calculation is diagonalizing the K matrix,
\begin{equation}
\label{eq-K-diag}
    K_{ii'} = \sum_{\rho} U_{i\rho} \mathrm{tan}(\tau_{\rho}) U^*_{i'\rho}\; ,
\end{equation}
to obtain the eigenphase shifts $\tau_{\rho}$ and the unitary (real and orthogonal in this case) matrix $U_{i\rho}$. The two energy dependent objects contain all the main information necessary for finding the potential curves and autoionizing states.

 The ability to evaluate Eq. (\ref{eq-Rmat-WE}) without the need to solve the Schr\"odinger equation again for each input energy is a key advantage of this approach. Eliminating the strongly closed channels while retaining the weakly closed ones in Eq. (\ref{eq-chan-elim-R}) is another key choice that gets rid of the more problematic channels that would otherwise hamper the evaluation of Eq. (\ref{eq-R2K}) while keeping the energy dependence of the resulting K matrix smoother than with a full elimination. Keeping some channels artificially open is also necessary for computing the potential curves below the lowest CH$^+$ target energy, where all channels are closed.

\subsection{\label{sec-theory-pot} Bound-state curve calculation}

This part of the approach is analogous to the multichannel spectroscopy treatment of bound-state energies by \citet{Aymar_Greene_LKoenig_1996} (subsection II.D.2 in the reference).

In this case, we are looking at energies below the lowest target threshold where all channels are closed, and thus we can only eliminate a strict subset of them. As noted earlier, we eliminate only the strongly closed channels because eliminating fewer channels introduces less energy dependence into the K matrix and its related quantities. Using the eigenphase shifts and their corresponding eigenvectors from Eq. (\ref{eq-K-diag}), we can construct two new matrices,
\begin{align}
\label{eq-sincos-diag}
    \mathcal{S}_{ii'} &= \sum_{\rho} U_{i\rho} \mathrm{sin}(\tau_{\rho}) U^*_{i'\rho}\; , \\
    \mathcal{C}_{ii'} &= \sum_{\rho} U_{i\rho} \mathrm{cos}(\tau_{\rho}) U^*_{i'\rho}\; ,
\end{align}
which can be seen as the coefficients of a unitary transformation of the standard real-valued asymptotic solution matrix $\psi_{ii'}$,
\begin{equation}
\label{eq-SC_solution}
\begin{gathered}
    \psi_{ii'} = f_i \delta_{ii'} - g_i K_{ii'}, \\
     \downarrow \\
    \Tilde{\psi}_{ii'} = f_i \mathcal{C}_{ii'} - g_i \mathcal{S}_{ii'}\;.
\end{gathered}
\end{equation}
They are to $\underline{K}$ what sine and cosine are to the tangent function.

Let us now write the Coulomb functions in terms of the energy normalized Whittaker functions 
\begin{align}
\label{eq-f_W_relation}
f_i &= - W_i \cos\beta_i + \bar{W}_i \sin\beta_i, \\
\label{eq-g_W_relation}
g_i &= - W_i \sin\beta_i - \bar{W}_i \cos\beta_i,
\end{align}
where $\beta_i=\pi(\nu_i-l_i)$.
Finding bound states means finding linear combinations of $\Tilde{\psi}_{ii'}$ such that the exponentially rising part is removed (when there are open channels, this exists for any energy). Using Eqs. (\ref{eq-SC_solution}), (\ref{eq-f_W_relation}), (\ref{eq-g_W_relation}) it can be seen that the bound-state energies can be found wherever the matrix
\begin{equation}
\label{eq-SCdet}
    \underline{D} = \underline{\mathcal{S}}\mathrm{cos}(\underline{\beta}) +
    \underline{\mathcal{C}}\mathrm{sin}(\underline{\beta}) \; ,
\end{equation}
is singular. Remembering that we sometimes replace $\{f,g\}$ with the rescaled $\{f^{\eta},g^{\eta}\}$, we need to adjust this to
\begin{equation}
\label{eq-SCdet-adjusted}
    \underline{\tilde{D}} = \underline{\mathcal{S}}\mathrm{cos}(\underline{\beta}) +
    \underline{\mathcal{C}}\mathrm{sin}(\underline{\beta})\underline{q}^{-1} \; ,
\end{equation}
where the matrix $\underline{q}$ is diagonal. Its elements are equal to 1 for channels that use $\{f,g\}$ and equal to $A(\epsilon_i,l_i)$ for channels that use $\{f^{\eta},g^{\eta}\}$.
Altogether, finding the potential curves is equivalent to finding the roots of the determinant
\begin{equation}
    \label{eq_Det}
    \left\vert\underline{\tilde{D}}\right\vert = 0
\end{equation}
with respect to energy for each fixed-nuclei calculation. The advantage of going from the K matrix to $\mathcal{S}$ and $\mathcal{C}$ is that $\mathcal{S}$ and $\mathcal{C}$ don't have divergences and are therefore easier to work with numerically.

\subsection{\label{sec-theory-autoion} Autoionizing curve calculation}

In the energy region between the target ground state and excited states lie the autoionizing curves. Their positions and widths are determined by the energy dependence of the eigenphase sum $\sum_{\rho}\tau_{\rho}(E)$ obtained from Eq. (\ref{eq-K-diag}) (see subsection II.D.3 of \citet{Aymar_Greene_LKoenig_1996}). The resonance positions are at the peaks of the eigenphase sum energy derivative and the full widths at half maximum are defined by
\begin{equation}
\label{eq-width}
    \Gamma = \frac{2}{\sum_{\rho} \left.\frac{d\tau_{\rho}}{dE} \right\vert_{E_\mathrm{peak}}} .
\end{equation}
\begin{figure}
  \centering
  \includegraphics[width=.48\textwidth]{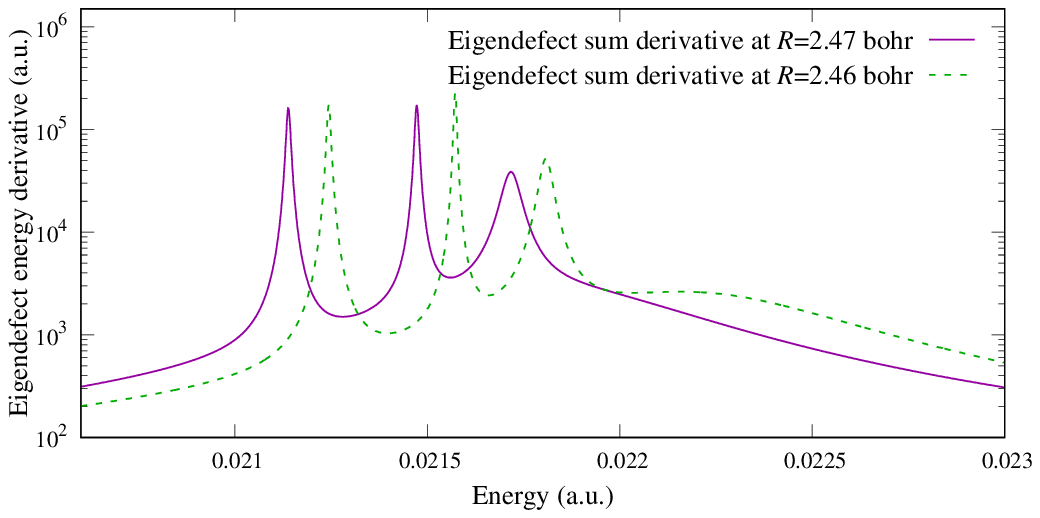}
  \caption{
    The energy derivative of the eigenphase sum at two inter-nuclear distances (for the $^2\Sigma^+$ symmetry, zoomed-in to a particular energy window). The widest peak visible at the $R=2.46$ bohr passes underneath the three narrow peaks at $R=2.47$ bohr.
  }
  \label{fig-sigP-peakloss}
\end{figure}
For this method to correctly determine the physical autoionization widths, it is necessary to eliminate all the closed channels before computing the eigenphase shifts. This leads to a sharper energy behavior of the K matrix compared to the method from the previous section, so in practice we must sample a denser energy grid.
A possible issue with this method is that peaks connected to wide resonances can sometimes be obfuscated in regions where they strongly overlap other narrower resonances. An example of this can be seen in Fig. \ref{fig-sigP-peakloss}.

A simpler, but less exact, alternative to this approach is to again keep some of the closed channels open when computing the K matrix, then cross out all the open channels and treat the remaining closed part with the procedure from subsection \ref{sec-theory-pot}. This simplified method neglects some of the channel couplings and does not allow determination of the resonance widths, but is useful as an approximation of the autoionizing curves because the resulting energies are usually very close to the ones obtained via the full approach (typically within one autoionization width $\Gamma$).

\subsection{\label{sec-theory-symmetry} Separating symmetries of the molecule}

The Quantemol software separates the problem into irreducible representations of the C$_\mathrm{2v}$ point group, namely $A_1$, $A_2$, $B_1$ and $B_2$. Hence, the R matrix is block diagonal with respect to these. After a simple unitary transformation of the R matrix, these blocks separate further into the actual physical symmetries of the problem. The transformation in question is connected to going from real-valued spherical harmonics to complex-valued ones, and is described briefly in our previous publication\cite{CHplus_DR}.
$A_1$ gives us $^2\Sigma^+$ and $^2\Delta$, $A_2$ gives $^2\Sigma^-$ and $^2\Delta$ (identical to $A_1$), $B_1$ gives $^2\Pi$ and $^2\Phi$ and $B_2$ holds the same information as $B_1$.

\section{\label{sec-results}Results and discussion}

\begin{figure}
  \centering
  \includegraphics[width=.48\textwidth]{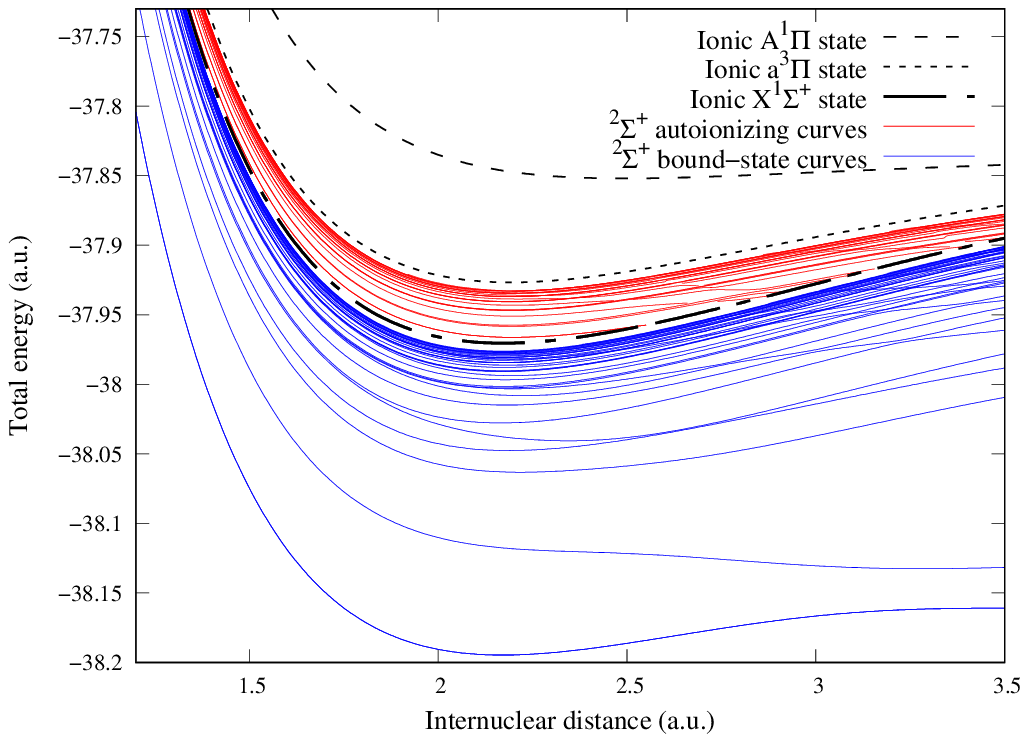}
  \caption{
    The full breadth of the $^2\Sigma^+$ symmetry curves of CH. The thick dashed curves are the CH$^+$ target states. The autoionizing curves are the continuous red curves between the ground target state and first excited target state. The bound-state curves are the continuous blue curves below the ground target state.
  }
  \label{fig-sigP-full}
\end{figure}

\begin{figure}
  \centering
  \includegraphics[width=.48\textwidth]{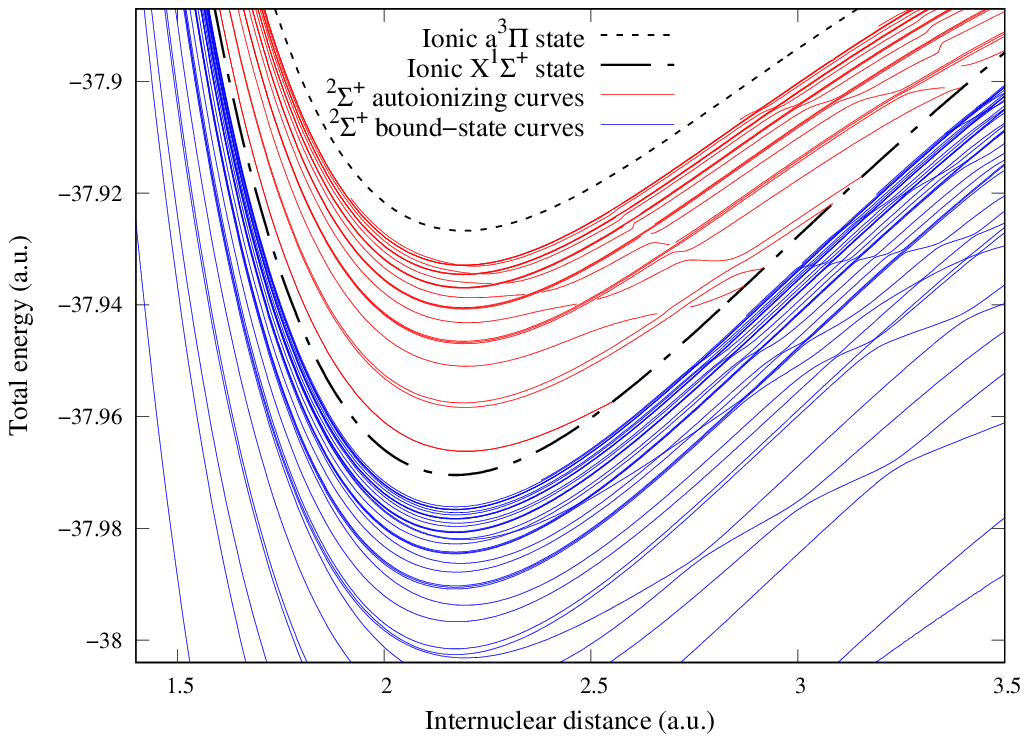}
  \caption{
    The $^2\Sigma^+$ symmetry curves zoomed in. The dashed curves are CH$^+$ target states. The continuous curves above X$^1\Sigma^+$ are autoionizing curves and the ones below are bound-state curves.
  }
  \label{fig-sigP-abs2}
\end{figure}

\begin{figure}
  \centering
  \includegraphics[width=.48\textwidth]{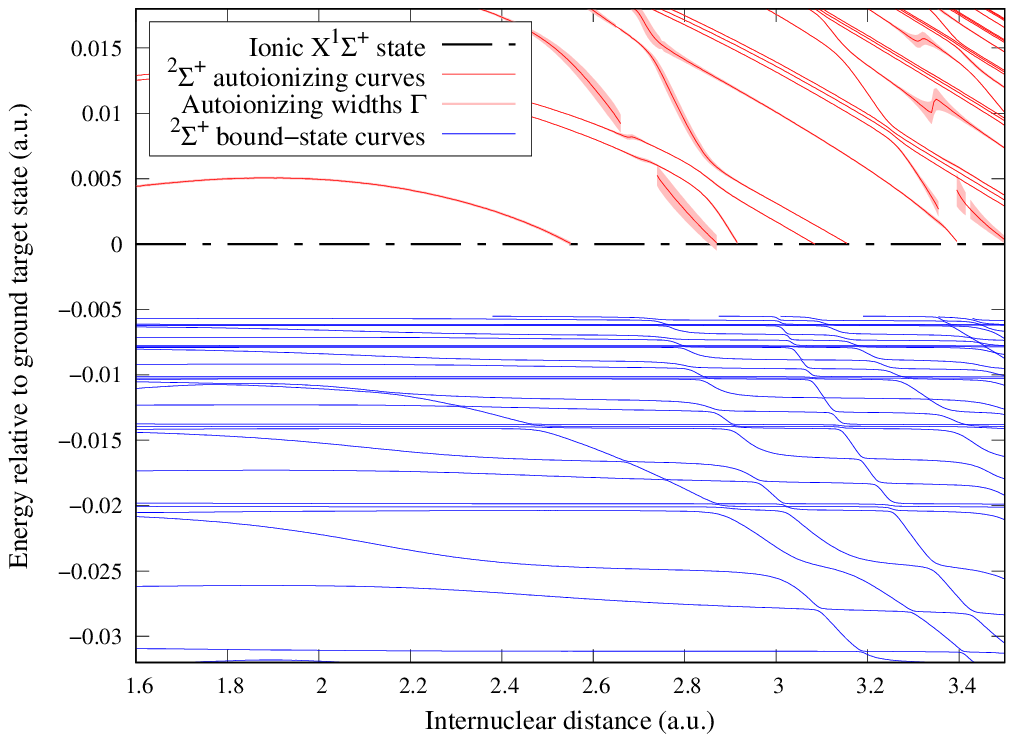}
  \caption{
    The $^2\Sigma^+$ symmetry curves relative to the ground target state energy (zoomed in to energies close to threshold). The autoionizing curves above threshold now also have a shaded area denoting curve widths.
  }
  \label{fig-sigP-rel2}
\end{figure}

\begin{figure}
  \centering
  \includegraphics[width=.48\textwidth]{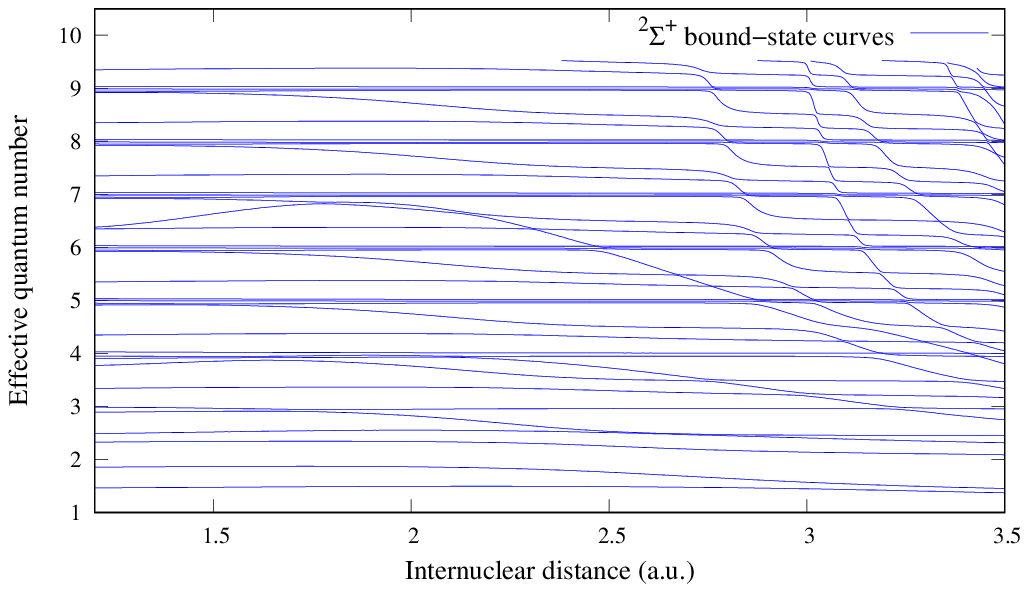}
  \caption{
    The $^2\Sigma^+$ symmetry bound-state curve effective quantum number (compared to ground target state).
  }
  \label{fig-sigP-nu}
\end{figure}

Because a Rydberg series gets infinitely dense in energy as it approaches the threshold, we cut off our calculations at 0.15 eV below each threshold.
The $^2\Sigma^+$ symmetry curves are in Figs.~\ref{fig-sigP-full}-~\ref{fig-sigP-nu}. Fig. \ref{fig-sigP-full} shows the overall structure over the full breadth of computed energies. Fig. \ref{fig-sigP-abs2} zooms in to energies closer around the ionic potential minimum. The structure is mostly dominated by a dense series of curves parallel to the closest target state above (ground state for the potential curves and first excited state for the autoionizing curves) which are \quotes{cut through} by a sparser series of plunging resonances. This creates many series of avoided crossings and is most apparent from Figs. \ref{fig-sigP-rel2} and \ref{fig-sigP-nu}, which show the curves relative to the ground state energy. Fig. \ref{fig-sigP-nu} exchanges energy for the effective quantum number $\nu$ (with respect to the ground state) which demonstrates how the majority of the potential curves are placed at integer $\nu$ positions (as expected of a Rydberg series).

\begin{figure}
  \centering
  \includegraphics[width=.48\textwidth]{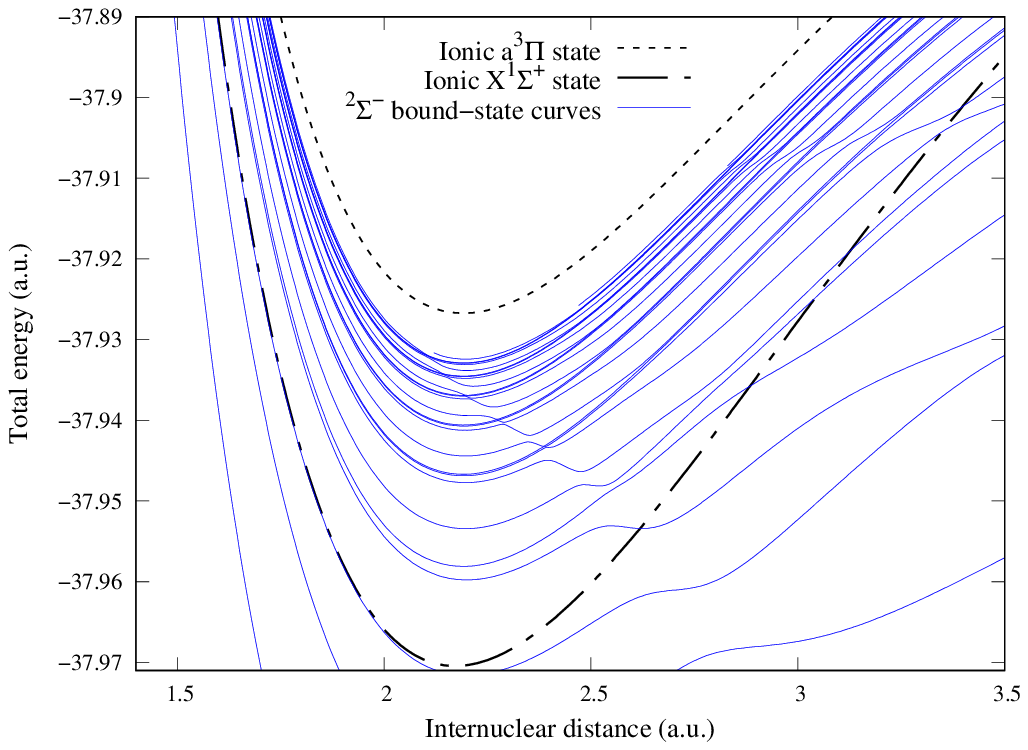}
  \caption{
    The $^2\Sigma^-$ symmetry potential curves. The dashed curves are CH$^+$ target states. All the continuous curves in this image are bound-state curves.
  }
  \label{fig-sigM-abs1}
\end{figure}

The remaining symmetries mostly feature similar behavior to that of $^2\Sigma^+$.
The $^2\Sigma^-$ symmetry potential curves are in Fig.~\ref{fig-sigM-abs1}. All the $^2\Sigma^-$ symmetry channels are connected to excited electronic states of the target, and thus accumulate under a higher threshold and don't feature any autoionizing curves in the studied energy regions.
\begin{figure}
  \centering
  \includegraphics[width=.48\textwidth]{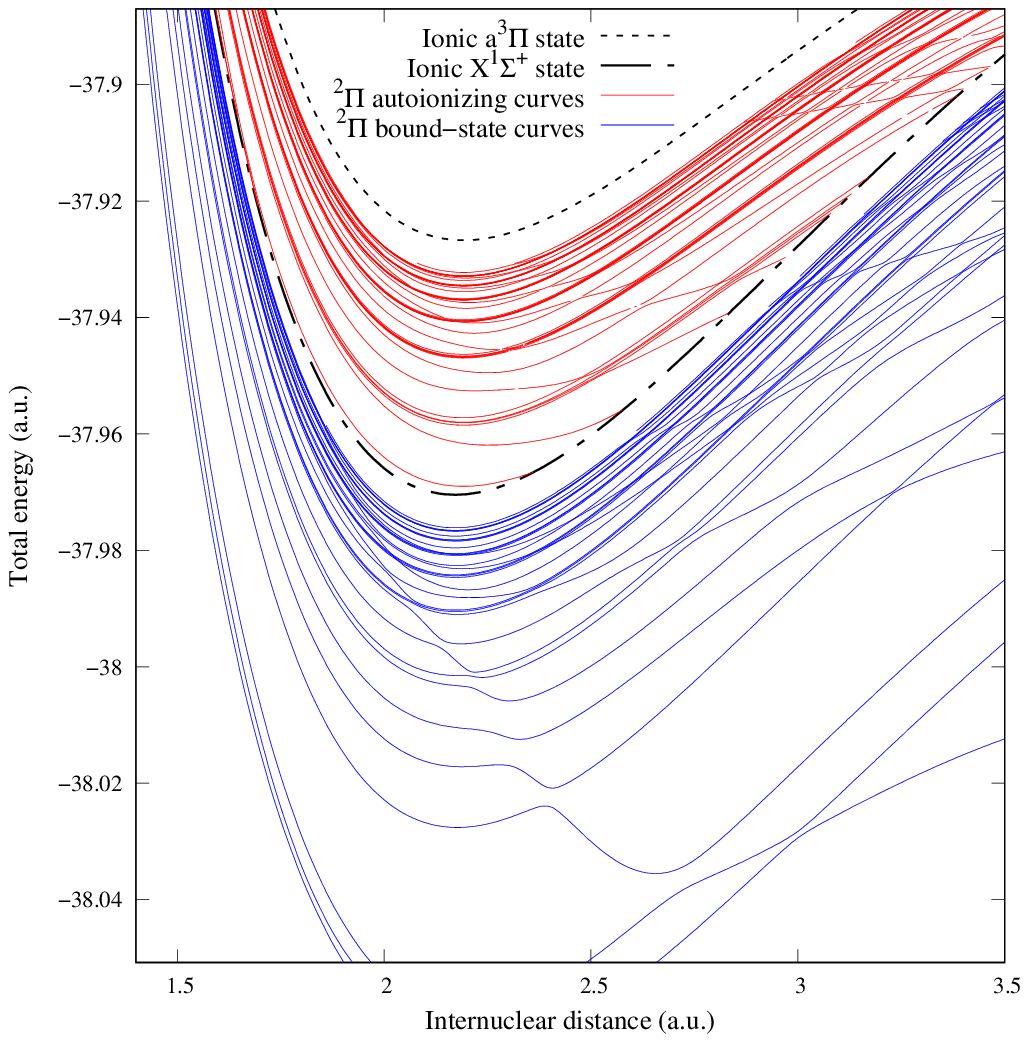}
  \caption{
    The $^2\Pi$ symmetry curves. The dashed curves are CH$^+$ target states. The continuous curves above X$^1\Sigma^+$ are autoionizing curves and the ones below are bound-state curves.
  }
  \label{fig-pi-abs1}
\end{figure}
\begin{figure}
  \centering
  \includegraphics[width=.48\textwidth]{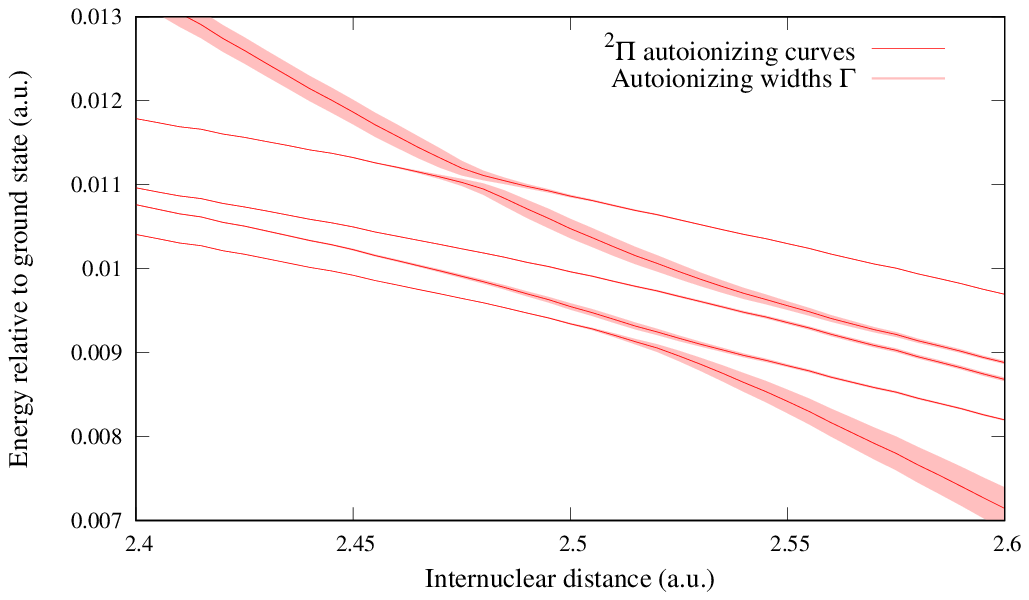}
  \caption{
    A closeup of a $^2\Pi$ symmetry autoionizing curve avoided crossing featuring the curve widths with shaded areas.
  }
  \label{fig-pi-avoided}
\end{figure}
\begin{figure}
  \centering
  \includegraphics[width=.48\textwidth]{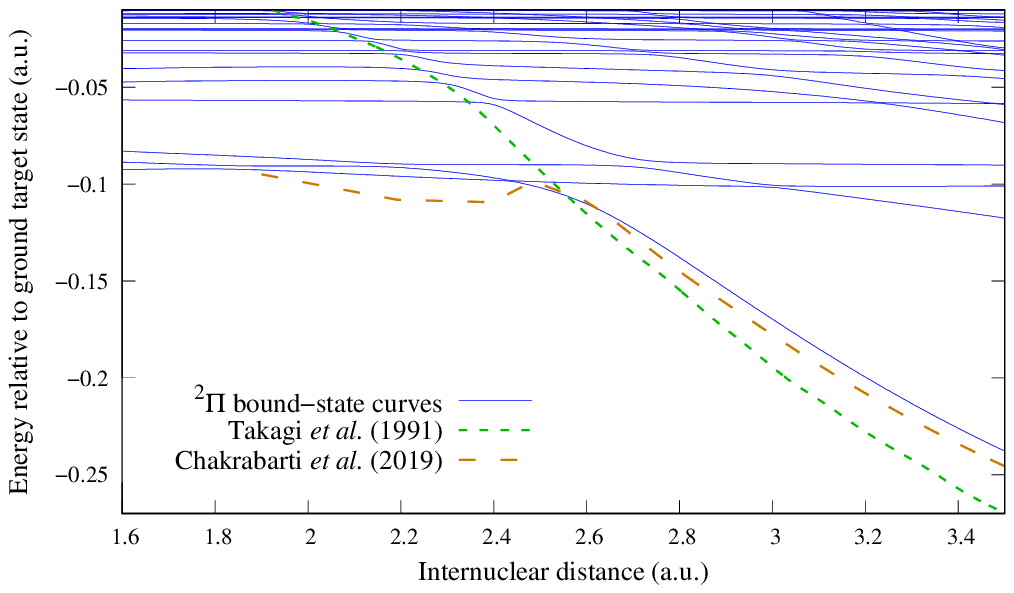}
  \caption{
    A comparison of the 2$^2\Pi$ curve from previous publications by \citet{takagi1991dissociative,chakrabarti2019r}. The continuous curves are our computed bound-state curves and the dashed curves are interpolated from figures in the aforementioned publications.
  }
  \label{fig-pi-comp}
\end{figure}
\begin{figure}
  \centering
  \includegraphics[width=.48\textwidth]{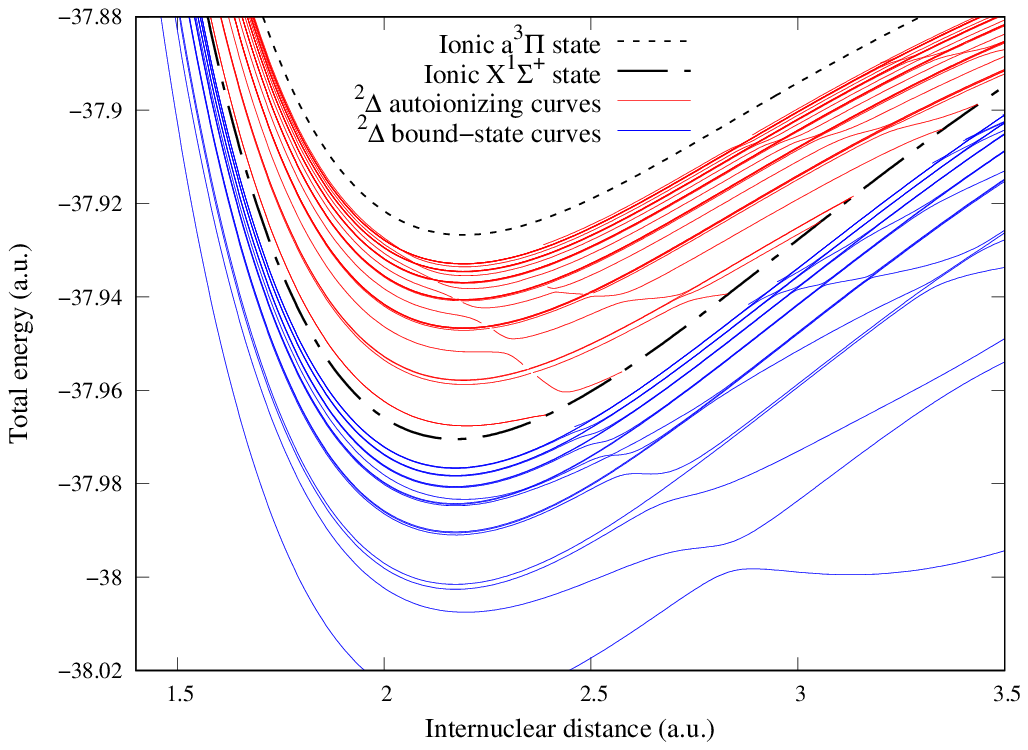}
  \caption{
    The $^2\Delta$ symmetry curves. The dashed curves are CH$^+$ target states. The continuous curves above X$^1\Sigma^+$ are autoionizing curves and the ones below are bound-state curves.
  }
  \label{fig-del-abs1}
\end{figure}
\begin{figure}
  \centering
  \includegraphics[width=.48\textwidth]{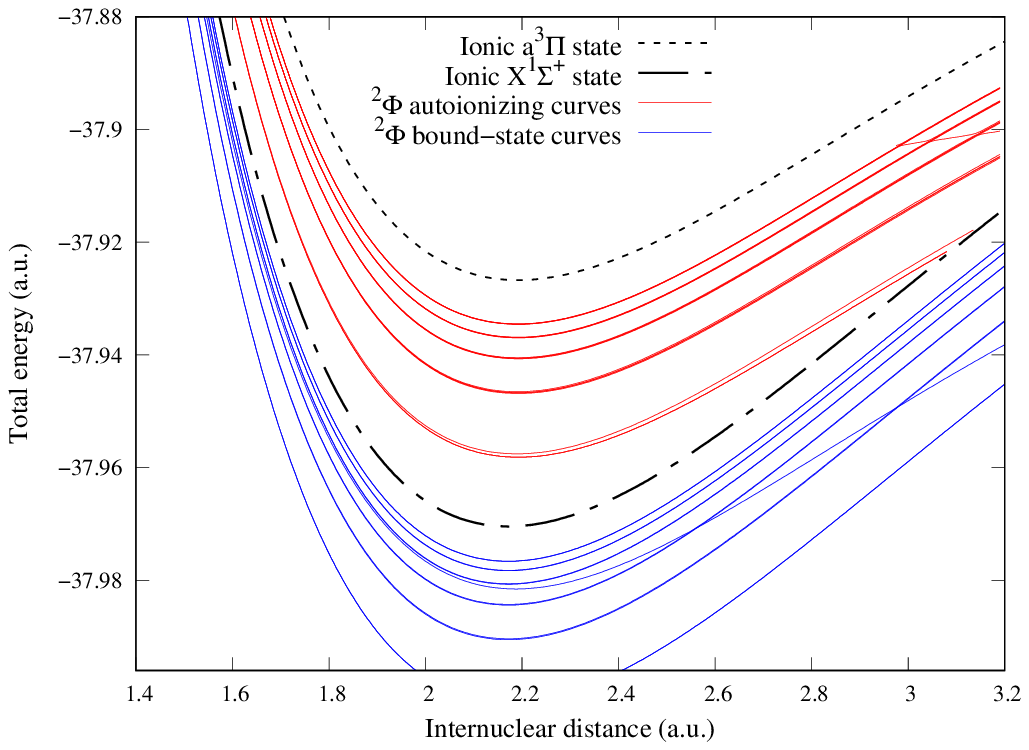}
  \caption{
    The $^2\Phi$ symmetry curves. The dashed curves are CH$^+$ target states. The continuous curves above X$^1\Sigma^+$ are autoionizing curves and the ones below are bound-state curves.
  }
  \label{fig-phi-abs1}
\end{figure}

The $^2\Pi$ symmetry curves are in Fig.~\ref{fig-pi-abs1} and feature some of the densest crossing structures. Fig.~\ref{fig-pi-avoided} shows a closeup of one of the avoided crossings including the computed widths $\Gamma$. It demonstrates how the largest width diabatically crosses through the curves. Often, though, these crossings are where we lose data points because the widest curve is often completely hidden behind the band of narrow curves.
Fig.~\ref{fig-pi-comp} shows a comparison with two previous calculations by \citet{takagi1991dissociative,chakrabarti2019r}. It shows the so-called 2$^2\Pi$ curve which is the lowest curve that cuts through the Rydberg series accumulating below X$^1\Sigma^+$. This curve is the continuation of the resonance which crosses X$^1\Sigma^+$ closest to its minimum. As such, it is the most relevant resonance to direct dissociative recombination computations such as those of \citet{Amitay_CHplus}. In our computations, the autoionizing curve cuts through X$^1\Sigma^+$ at 1.7 bohr. Do note that the compared curves have been digitized and interpolated from the aforementioned articles and in the case of Chakrabarti \textit{et al.} said curve only has four data points in the region from 2-2.5 bohr.

The $^2\Delta$ symmetry potential curves are in Fig.~\ref{fig-del-abs1}.
The $^2\Phi$ symmetry potential curves, on the other hand, are almost exclusively at integer $\nu$ energies (relative to either the ground or first excited target state); they are in Fig.~\ref{fig-phi-abs1}.

\subsection{\label{sec-results-limitations}Numerical limitations}

In general, the widths of autoionizing curves are proportional to the Rydberg spacing~\cite{fano_rau_book}
\begin{equation}
\label{eq-width2}
    \Gamma(\epsilon) \propto \frac{\mathrm{d}\epsilon}{\mathrm{d}\nu}(\epsilon)=\frac{1}{\nu^3(\epsilon)},
\end{equation}
where $\epsilon$ is computed relative to the threshold that makes up the dominant channels of a given curve. The actual curve widths can vary significantly around the avoided crossings but within simpler regions (e.g. at internuclear distances less than 2 bohr for $^2\Sigma^+$) this scaling rule is mostly followed. If we compute the ratios $\Gamma_\text{curve}\nu^3(\epsilon_\text{curve})$ for all the curves, they form tight bands with a variance of approximately 5-15 percent.
In practice, this scaling means that one should sample the energy grid with equidistant $\nu$ (relative to the closest higher threshold). For our specific computational setup, this means that widths narrower than
\begin{equation}
\label{eq-width-cutoff}
    \Gamma_{\mathrm{limit}}(\epsilon) = \frac{8.639 \times 10^{-4}}{\nu^3(\epsilon)}\mathrm{a.u.}
\end{equation}
start becoming unreliable and curves with significantly lower widths are likely to be lost ($\epsilon$ is again relative to the closest higher threshold).
In particular, the $^2\Phi$ symmetry is almost completely made up of curves that skirt this boundary.

\section{\label{sec-conclusions}Conclusions}
We have processed fixed-nuclei R matrix data from Quantemol using two methods to obtain bound-state curves and autoionizing curves in the energy region around the X$^1\Sigma^+$ state of CH$^+$. The computations were done for five symmetries of the molecule: $^2\Sigma^+$, $^2\Sigma^-$, $^2\Pi$, $^2\Delta$, and $^2\Phi$. The autoionizing curve data also features computed autoionizing widths.

\section{\label{sec-supplement}Supplementary material description}
The supplement to this article contains data files for all the curves shown in Figs. ~\ref{fig-sigP-full}-~\ref{fig-phi-abs1} (apart from the two curves extracted from previous studies in Fig.\ref{fig-pi-comp}). The files are separated by symmetry and further separated into autoionizing curve files and bound-state curve files. All values are in atomic units.

In all of the files the first column contains the internuclear distance and the second, third and fourth columns contain the X$^1\Sigma^+$, a$^3\Pi$ and A$^1\Pi$ energies respectively. For the autoionizing curves, the following columns come in pairs containing a curve's energy relative to X$^1\Sigma^+$ followed by its width $\Gamma$. The $^2\Phi$ symmetry autoionizing curve file does not contain widths due to insufficient resolution. The bound-state curve files also don't contain any widths. In cases where the data for a curve doesn't exist at a specific geometry (either because it isn't in the studied energy range or because it was temporarily lost in an avoided crossing as described in Fig.~\ref{fig-sigP-peakloss}) the data entry at that position is a "?" symbol. The set of geometries used is the same for all files.

\begin{acknowledgments}
We wish to acknowledge support from the National Science Foundation, Grant Nos.2110279 (UCF) and 2102187 (Purdue).
\end{acknowledgments}

\section*{Author declarations}
The authors have no conflicts to disclose

\section*{Data Availability Statement}

The data that support the findings of this study are available within the article and its supplementary material.


\appendix*

\section{\label{sec-R_elim}R-matrix preliminary channel elimination}

Because R-matrix channel elimination is not very common, we find it useful to write up our derivation for the formula (\ref{eq-chan-elim-R}).
The ideas used here are similar to Gregory Breit's reduced R-matrix treatment\cite{Breit_1959} (section 31 in the reference). We call this a preliminary elimination because in our frame transformation treatments\cite{CHplus_DR} we use it to eliminate only the strongly closed channels in order to obtain a smooth fixed-nuclei K matrix and it is the rovibrationally frame transformed K matrix that undergoes the full MQDT closed-channel elimination.

As a first step, recall that Eq. (\ref{eq-R2K}) can be derived from the definition of the R matrix as an inverse logarithmic derivative of a wave function. Let $\psi_{i\gamma}$ be the $i^\text{th}$ channel of the $\gamma^\text{th}$ independent solution of the Schr\"odinger equation. In matrix formalism, one can write
\begin{equation}
    \underline{R} = \left. \underline{\psi}(\underline{\psi}')^{-1} \right\vert_{r_0} ,
\end{equation}
where $r_0$ is the R-matrix box radius. This equation is valid for any choice of independent solutions. This means that for $r>r_0$, we can substitute the standard Coulomb functions $\{f,g\}$ and the matrix $\underline{K}$ with an arbitrary pair of solutions to the radial Schr\"odinger equation $\{\Tilde{f},\Tilde{g}\}$ and their corresponding combination matrix $\underline{\Tilde{K}}$, i.e.
\begin{equation}
    \underline{\psi}(r)=\underline{\tilde{f}}-\underline{\tilde{g}}\underline{\tilde{K}}, \quad \text{for }r\geq r_0.
\end{equation}
If we choose such a pair that satisfies the conditions on the R-matrix box boundary
\begin{equation}
\label{eq-fg_bcond}
\begin{aligned}
    \Tilde{f}(\epsilon,l,r_0) = 0, \quad \Tilde{f}'(\epsilon,l,r_0) = 1, \\
    \Tilde{g}(\epsilon,l,r_0) = 1, \quad \Tilde{g}'(\epsilon,l,r_0) = 0,
\end{aligned}
\end{equation}
then Eq. (\ref{eq-R2K}) becomes $\underline{\Tilde{K}}=-\underline{R}$ (at $r=r_0$). This means that we can also view the R matrix as a K matrix.
Let us specifically choose the set of Coulomb functions
\begin{equation}
\label{eq-fgW}
\begin{aligned}
    \Tilde{f}(r) &= -\frac{\pi}{2}\left[ W(r)\bar{W}(r_0) - \bar{W}(r)W(r_0) \right], \\
    \Tilde{f}'(r) &= -\frac{\pi}{2}\left[ W'(r)\bar{W}(r_0) - \bar{W}'(r)W(r_0) \right], \\
    \Tilde{g}(r) &= \phantom{-}\frac{\pi}{2}\left[ W(r)\bar{W}'(r_0) - \bar{W}(r)W'(r_0) \right], \\
    \Tilde{g}'(r) &= \phantom{-}\frac{\pi}{2}\left[ W'(r)\bar{W}'(r_0) - \bar{W}'(r)W'(r_0) \right],
\end{aligned}
\end{equation}
where $W$ and $\bar{W}$ are a pair of \quotes{energy normalized} Whittaker functions\cite{Aymar_Greene_LKoenig_1996} with $W$ exponentially decaying for closed channels and $\bar{W}$ exponentially increasing. The $\epsilon$ and $l$ arguments are omitted for brevity.

Keeping with the standard MQDT closed-channel elimination procedure\cite{Aymar_Greene_LKoenig_1996}, we first partition the solution matrix into open and closed blocks
\begin{equation}
        \underline{\psi}=\begin{pmatrix}
        \underline{\psi_{oo}} & \underline{\psi_{oc}}\\ 
        \underline{\psi_{co}} & \underline{\psi_{cc}}
\end{pmatrix},
\end{equation}
and then look for a linear combination matrix $\underline{B}$ that kills off the exponentially rising behavior in the closed channels.
\begin{equation}
\label{eq-fg_lin_comb}
        \underline{\psi}^{\mathrm{phys}}=\underline{\psi}\begin{pmatrix}
        \underline{B_{oo}} \\ 
        \underline{B_{co}} \end{pmatrix}
        \rightarrow \begin{pmatrix}
        [\Tilde{f}_o-\tilde{g}_o\underline{\Tilde{K}_{oo}}]\underline{B_{oo}} + [-\tilde{g}_o\underline{\Tilde{K}_{oc}}]\underline{B_{co}}  \\ 
        [-\tilde{g}_c\underline{\Tilde{K}_{co}}]\underline{B_{oo}} +[\Tilde{f}_c-\tilde{g}_c\underline{\Tilde{K}_{cc}}]\underline{B_{co}}
\end{pmatrix}.
\end{equation}
Combining Eqs. (\ref{eq-fgW}) and (\ref{eq-fg_lin_comb}) while requiring that coefficients in front of $\bar{W}$ be zero in all closed channels and evaluating at $r=r_0$ yields
\begin{equation}
    \underline{B_{co}} = \left(\underline{W_c}(r_0) + \underline{W_c'}(r_0)\underline{\tilde{K}_{cc}} \right)^{-1} \left( -\underline{W_c'}(r_0) \underline{\tilde{K}_{co}} \right) \underline{B_{oo}}.
\end{equation}
Setting the arbitrary $B_{oo}$ equal to the identity matrix and substituting it into the open part of Eq. (\ref{eq-fg_lin_comb}), $\underline{\Tilde{K}}=-\underline{R}$ then gives us 
\begin{equation}
    \label{eq-CE_R}
    \underline{R}^\mathrm{phys} = \underline{R_{oo}} - \underline{R_{oc}}
    \left( \underline{R_{cc}} - \underline{W_c}\left[\underline{W'_c}\right]^{-1} \right)^{-1} \underline{R_{co}}\; .
\end{equation}
Note here that an $r$-independent normalization of $W$ does not change this equation.

\section{\label{sec-Moredata}Additional curve properties}

\begin{figure}
  \centering
  \includegraphics[width=.48\textwidth]{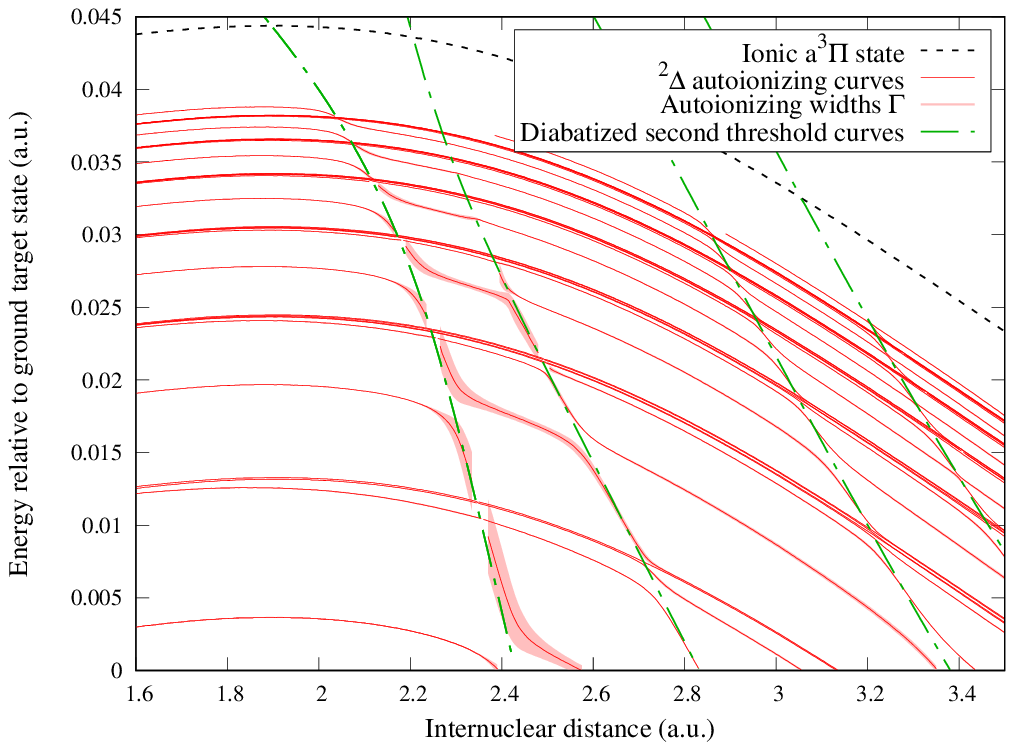}
  \caption{
    The $^2\Delta$ symmetry autoionizing curves (continuous red curves with shaded width) compared to diabatized curves (dot-dashed green curves).
  }
  \label{fig-del-diab}
\end{figure}
\begin{figure}
  \centering
  \includegraphics[width=.48\textwidth]{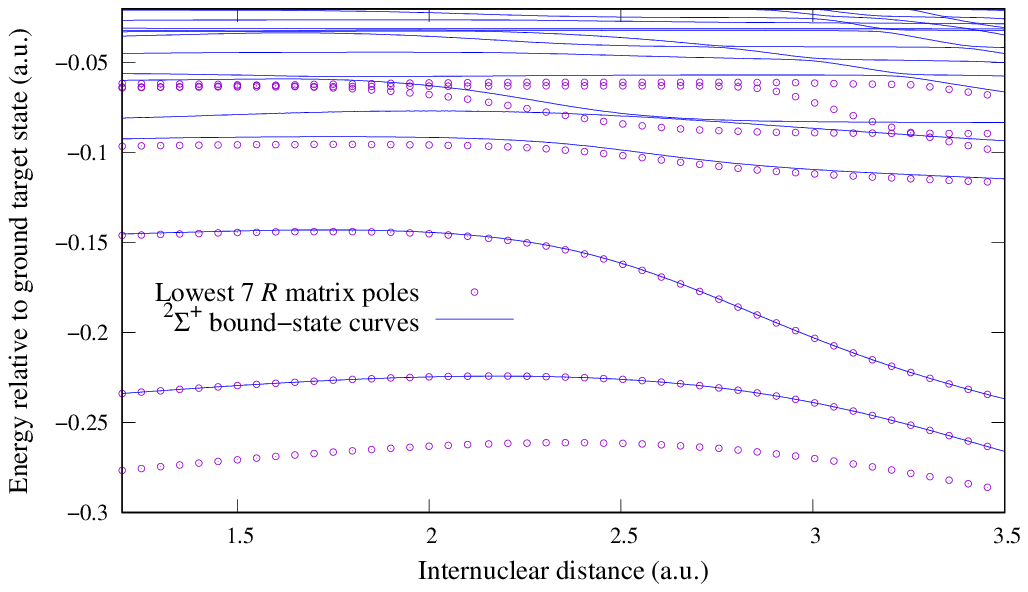}
  \caption{
    The $^2\Sigma^+$ symmetry bound-state curves relative to the ground target state energy showing that the lowest curves coincide with the low R-matrix poles.
  }
  \label{fig-sigP-rel1}
\end{figure}

If we artificially open all the channels in our K matrix connected to the first excited threshold and still implement the autoionizing curve procedure, we will get an approximate diabatization of the plunging resonances connected to the higher threshold, as shown in Fig.~\ref{fig-del-diab}.

Another property of fixed-nuclei R-matrix data is that the lowest-lying bound-state curves coincide with the lowest-lying R matrix poles. This is shown in Fig.~\ref{fig-sigP-rel1}. The one set of poles that does not coincide with any curve shown, instead coincides with the lowest curve in the $^2\Delta$ symmetry.

\bibliography{REFS.bib}

\end{document}